\documentstyle[12pt]{article}

\catcode`\@=11
\long\def\@makefntext#1{
\protect\noindent \hbox to 3.2pt {\hskip-.9pt
$^{{\ninerm\@thefnmark}}$\hfil}#1\hfill}                

 \def\@makefnmark{\hbox to 0pt{$^{\@thefnmark}$\hss}}  
 
\def\ps@myheadings{\let\@mkboth\@gobbletwo
\def\@oddhead{\hbox{}
\rightmark\hfil\ninerm\thepage}
\def\@oddfoot{}\def\@evenhead{\ninerm\thepage\hfil
\leftmark\hbox{}}\def\@evenfoot{}
\def\sectionmark##1{}\def\subsectionmark##1{}}
 
\newcounter{sectionc}\newcounter{subsectionc}\newcounter{subsubsectionc}
\renewcommand{\section}[1] {\vspace{0.6cm}\addtocounter{sectionc}{1}
\setcounter{subsectionc}{0}\setcounter{subsubsectionc}{0}\noindent
	{\bf\thesectionc. #1}\par\vspace{0.4cm}}
\renewcommand{\subsection}[1] {\vspace{0.6cm}\addtocounter{subsectionc}{1}
	\setcounter{subsubsectionc}{0}\noindent
	{\it\thesectionc.\thesubsectionc. #1}\par\vspace{0.4cm}}
\renewcommand{\subsubsection}[1] {\vspace{0.6cm}\addtocounter{subsubsectionc}{1}
	\noindent {\rm\thesectionc.\thesubsectionc.\thesubsubsectionc.
	#1}\par\vspace{0.4cm}}

\newcounter{appendixc}
\newcounter{subappendixc}[appendixc]
\newcounter{subsubappendixc}[subappendixc]

\renewcommand{\appendix}[1] {\vspace{0.6cm}
	\refstepcounter{appendixc}
	\setcounter{figure}{0}
	\setcounter{table}{0}
	\setcounter{equation}{0}
	\renewcommand{\thefigure}{\Alph{appendixc}.\arabic{figure}}
	\renewcommand{\thetable}{\Alph{appendixc}.\arabic{table}}
	\renewcommand{\theappendixc}{\Alph{appendixc}}
	\renewcommand{\theequation}{\Alph{appendixc}.\arabic{equation}}
	\noindent{\bf Appendix \theappendixc #1}\par\vspace{0.4cm}}

\renewenvironment{thebibliography}[1]
	{\begin{list}{\arabic{enumi}.}
	{\usecounter{enumi}\setlength{\parsep}{0pt}
\setlength{\leftmargin 1.25cm}{\rightmargin 0pt}
	 \setlength{\itemsep}{0pt} \settowidth
	{\labelwidth}{#1.}\sloppy}}{\end{list}}
 
\newcounter{itemlistc}
\newcounter{romanlistc}
\newcounter{alphlistc}
\newcounter{arabiclistc}

\newcommand{\fcaption}[1]{
	\refstepcounter{figure}
	\setbox\@tempboxa = \hbox{\tenrm Fig.~\thefigure. #1}
	\ifdim \wd\@tempboxa > 6in
	   {\begin{center}
	\parbox{6in}{\tenrm\baselineskip=12pt Fig.~\thefigure. #1}
	    \end{center}}
	\else
	     {\begin{center}
	     {\tenrm Fig.~\thefigure. #1}
	      \end{center}}
	\fi}
 
\newcommand{\tcaption}[1]{
	\refstepcounter{table}
	\setbox\@tempboxa = \hbox{\tenrm Table~\thetable. #1}
	\ifdim \wd\@tempboxa > 6in
	   {\begin{center}
	\parbox{6in}{\tenrm\baselineskip=12pt Table~\thetable. #1}
	    \end{center}}
	\else
	     {\begin{center}
	     {\tenrm Table~\thetable. #1}
	      \end{center}}
	\fi}
 
\def\@citex[#1]#2{\if@filesw\immediate\write\@auxout
	{\string\citation{#2}}\fi
\def\@citea{}\@cite{\@for\@citeb:=#2\do
	{\@citea\def\@citea{,}\@ifundefined
	{b@\@citeb}{{\bf ?}\@warning
	{Citation `\@citeb' on page \thepage \space undefined}}
	{\csname b@\@citeb\endcsname}}}{#1}}
 
\newif\if@cghi
\def\cite{\@cghitrue\@ifnextchar [{\@tempswatrue
	\@citex}{\@tempswafalse\@citex[]}}
\def\citelow{\@cghifalse\@ifnextchar [{\@tempswatrue
	\@citex}{\@tempswafalse\@citex[]}}
\def\@cite#1#2{{$\null^{#1}$\if@tempswa\typeout
	{IJCGA warning: optional citation argument
	ignored: `#2'} \fi}}

\def\fnt#1#2{\footnotetext{\kern-.3em
	{$^{\mbox{\sevenrm #1}}$}{#2}}}
 
 1
 1
 1

\font\tenrm=cmr10

\font\ninerm=cmr9

\begin{document}

\vspace{10 mm}
\begin{center}
\large{{\bf The "Pioneer effect" as a manifestation of the cosmic expansion
in the solar system}}
\end{center}

\vspace{3 mm}
\hspace{2 cm}  J.L. Rosales \footnote{E-mail: rosales@delta.ft.uam.es}

\hspace{2 cm}
     {\em Departamento de F\'isica Moderna, Facultad de Ciencias }

\hspace{2 cm} {\em Universidad de Cantabria, 39005, Santander, Spain.}

\vspace{4 mm}

\hspace{2 cm}  J.L. S\'anchez-Gomez \footnote{E-mail: jolu@delta.ft.uam.es}

\hspace{2 cm}
     {\em Departamento de F\'isica Te\'orica, Facultad de Ciencias }

\hspace{2 cm} {\em Universidad Aut\'ooma de Madrid, 28049, Canto Blanco, Madrid, 
Spain.}

\vspace{3 mm}

\begin{quote}
\begin{center}
				Abstract
\end{center}

It is proposed that the recently reported  anomalous acceleration 
acting on the Pioneers spacecrafts should be a consequence of the existence of 
some local curvature in light geodesics when using the coordinate speed of
light in an expanding spacetime. This
suggests that the Pioneer effect is 
nothing else but the detection of cosmological expansion in the solar system.

\end{quote}
\vspace{2 mm}

{\it Pioneer effect.} A careful analysis of orbital data from  Pioneer
10/11 spacecrafts has been reported\cite{kn:Pioneer} which indicates 
the existence  of a very weak acceleration - approximately
$\kappa\simeq 8.5 \cdot 10^{-8}cm/s^2$ - directed
toward the Sun. The most conservative (or less adventurous)
hypothesis is that the Pioneer effect  does not entail new 
physics and that the detected misfit must be due to  some 
sophisticated (technological) reason 
having to do with  the spacecraft configuration.
However, the analysis in\cite{kn:Pioneer} seems to have ruled out many 
(perhaps all) of such technical
reasons  and the authors even claim  having 
taken into account the accepted
values of the errors in the planetary ephemeris, Earth's orientation, precesion,
and nutation.

Thus, in principle, a new effect seems to have unexpectly 
entered the  phenomenology of physics. 
On the other hand, if such an effect really exists
(i.e., it can not be eliminated by  data reanalysis) it would represent
a violation of Birkhoff's theorem in general relativity for no
constant acceleration at all is  predicted by the Schwarzschild solution.
The nature of the effect is still far from being clarified. One of its 
surprising features, as pointed out in \cite{kn:Pioneer}, is that it does not
affect the planets (since no cumulative precession is observed in 
their trajectories) but only to objects with masses similar to that of 
spacecrafts (an apparently, strong violation of the equivalence principle!)

Apart from their masses, the only sensible difference between planets and
spacecrafts is the nature of their corresponding orbits, i.e., their relative
motion with respect to the Sun. That is why the effect might be
originated from some unexpected correction to the way we compute relative
motions in the solar system.

An aspect that has not been analyzed in planetary ephemeris is the observed
difference of the expansion of the universe for nearby points (the Sun
and the planets, say). This could amount to correcting the relative
accelerations as computed as a given time of cosmological expansion.

Let us start by considering  trajectories in a FRW metric
\begin{equation}
ds^2=-c^2dt^2+\chi(t)^2 dr^2 \mbox{\hspace{2 mm},}
\end{equation}
where, taking our units of space and time at the cosmological time $t_1$,
we can write,
\begin{equation}
\chi(t)\simeq (\frac{t}{t_1})^p \mbox{\hspace{2 mm},}
\end{equation}
where $p<1$  is a constant depending on the density of the universe, 
and $t_1$ is the local "cosmic time".

Light geodesics satisfy
\begin{equation}
dl\equiv cdt=\chi dr            \mbox{\hspace{2 mm},}
\end{equation}
where $dl$ is the lenght on the null cone. 

On the other hand, 
we should be able to write our physical laws in such a way that the expansion
of space time be scaled out. This requires using the radial function,
\begin{equation}
r_*\equiv \chi r \mbox{\hspace{2 mm},}
\end{equation}
the metric then becoming
\begin{equation}
ds^2=-(1-\frac{r_*^2H^2}{c^2})c^2dt^2+dr_*^2-2r_*Hdr_*dt \mbox{\hspace{2 mm}.}
\end{equation}
where the local Hubble parameter is
\begin{equation}
H=\frac{d}{dt}\log(\chi) \mbox{\hspace{2 mm}.}
\end{equation}
Since these are not syncronous coordinates (for $g_{0r_*}\neq 0$), 
we define the radial vector
\begin{equation}
\vec{g}=\frac{r_* H/c}{1-r_*^2H^2/c^2}\vec{r}_1 \mbox{\hspace{2 mm},}
\end{equation}
so that the  space like  element, as measured by some local observer, 
is  the embedded three dimensional metric within the global space time in 
this manifold (see e.g. \cite{kn:Lichnerowicz}) 
\begin{equation}
dl_*^2=(g_{r_* r_*}-g_{00}g^2)dr_*^2=\frac{dr_*^2}{1-r_*^2H^2/c^2} \mbox{\hspace{2 mm}.}
\end{equation}

Me may now compare the lenght, $l_*$ in the locally scaled coordinates, with 
$l$ on the light cone.
In order to do this, notice that one might also have obtained, after (3) and (4), 
the following equation  for the null cone,
\begin{equation}
dl=dr_*(l_*)-r_*(l_*)H\frac{dl}{c} \mbox{\hspace{2 mm},}
\end{equation}
whose solution - using (8), and noting that $\dot{H}\sim O(H^2)$- is
\begin{equation}
l=\frac{c}{H}\log\{1+\sin(\frac{Hl_*}{c})\}\simeq l_*-\frac{Hl_*^2}{2c}+O(H^2) \mbox{\hspace{2 mm}.}
\end{equation}

This represents the measure of the space time curvature on the 
local null cone.

Now, let us return to the original problem of the detected misfit between
the calculated and the measured position in the spacecrafts. First, we must
take into account that Sun's gravitational interaction takes place on points
on the light cone. This means that the real position of any particle with
respect to the center of forces is $l$, and, after Equation (10), we
realize that this  quantity differs from the expected distance, $ l_*$, 
(as computed upon using the local frame)
just in a systematic "bias" consisting on an effective residual acceleration  directed
toward the center of coordinates; its constant value is
\begin{equation}
\kappa=Hc \mbox{\hspace{2 mm}.}
\end{equation}
This is the   acceleration observed in Pioneer 10/11
spacecrafts. From the value reported in \cite{kn:Pioneer} we get a  value for 
Hubble parameter
\begin{equation}
H=\frac{\kappa}{c}\simeq 85 km/ s \cdot Mpc \mbox{\hspace{2 mm}.}
\end{equation}

The remarkable fact is that (10) is a function of the radius not of the time.
This means that a periodic orbit does not experience the systematic bias
but only a very small correction
\begin{equation}
-\frac{H}{2c}l_*^2\sim 10^{-4} \mbox{meters\hspace{2 mm},}
\end{equation}
which is not detectable.

The observed fall to the center of forces does not
entail cumulative effects in the orbital parameters since the result is
frame dependent. This is easily seen. As stated above, the parameter $p$ is
a function of the density of the Universe, $\mu$. Let us here make the 
approximation $\mu\simeq 0$, i.e., $p=1-\delta$, $\delta\ll 1$,
-recall that $\delta$ is proportional to the global density of the universe and we
must neglect its contribution in comparation with the local density
of matter in the solar system. 
Since, on the other hand, the curvature of the light cone only depends upon the time 
development of the manifold, then, in order to prevent the existence of
observable cumulative precesion on the orbits, we require
some new set of space and {\it time}  coordinates. We can select, for instance, 
the following transformation which, for $p=1$, relates the Lorentzian metric
with the Milne ($\chi=Ht$) space time:

\begin{equation}
dt=\frac{1}{(c^2\tau^2-\tilde{r}^2)^{1/2}}[c\tau d\tau -\frac{\tilde{r}}{c}d\tilde{r}] \mbox{\hspace{2 mm},}
\end{equation}
\begin{equation}
dr=\frac{pc^2}{H(c^2\tau^2-\tilde{r}^2)}[\tau d\tilde{r}-\tilde{r}d\tau] \mbox{\hspace{2 mm}.}
\end{equation}
Using these transformations,  Equation (1) becomes
\begin{equation}
ds^2 \rightarrow -c^2 d\tau^2 + d\tilde{r}^2 + \frac{2c^2 \delta}{\tilde{r}^2-c^2\tau^2}\log[eH(\tau^2-\frac{\tilde{r}^2}{c^2})^{1/2}]\{\tilde{r}d\tau-\tau d\tilde{r}\}^2      
\mbox{\hspace{2 mm}.}
\end{equation}
For $\delta\simeq 0$, it corresponds to the Minkowski space 
Now, $ds^2=0$ leads to the differential equation
\begin{equation}
\frac{d\tilde{r}}{d\tau}\simeq c\{1+\delta\frac{c\tau-\tilde{r}}{c\tau+\tilde{r}}\log[eH(\tau^2-\frac{\tilde{r}^2}{c^2})^{1/2}]\} \mbox{\hspace{2 mm},}
\end{equation}
whose solution is fairly simple, $\tilde{r}=c\tau + O(H^2)$ - independently of the value 
of $H$.
This means that the effect can be exactly removed in these coordinates.
This satisfactory fact agrees with the consequences of Birkhoff's theorem.
\vspace{6 mm}

{\it Conclusions}. The special features of the physical phenomena, 
including those properties that correspond
to the motion of the bodies, become
different in different systems of coordinates. This fact is illustrated,
for instance, in the old Foucault pendulum experiment. There, the motion 
of the pendulum experiences the effect of the Earth based reference system
-being not an inertial frame relatively to the "distant
stars". We have learnt, from the previous arguments, that {\it Pioneer effect} is
a kind of a new cosmological Foucault experiment, the solar system based
coordinates,  being not  the true  inertial frame with respect to the expansion of the 
universe, mimics the role that 
the rotating Earth  plays in Foucault's experiment.

\vspace{6 mm}
{\it Acknowledgements.} This work has been partialy supported by 
C. I. C. y T. (Spain) under contract PB-95-594.
J.L.R. wishes to thank to the Departamento de F\'isica Te\'orica 
of the Universidad Aut\'onoma de Madrid for  kind hospitality during the
completion of this work.


\begin{thebibliography}{99}
\bibitem{kn:Pioneer} J.D. Anderson et al. {\em Phys. Rev. Lett.} {\bf 81}, 2858, (1998).

\bibitem{kn:Lichnerowicz} A. Lichnerowicz, {\em Th. Rel. de la Grav. et de
l' Elec.,} Masson (1955).





\end{thebibliography}
\end{document}